\documentstyle[amsfonts,12pt]{article}

\begin{document}

\def\nqq{\hspace{-2em}}
\def\beq#1{\begin{equation}\label{#1}}
\def\eeq{\end{equation}}
\def\ber#1{\begin{eqnarray}\label{#1} \nqq}
\def\eer{\end{eqnarray}}
\def\nn{\nonumber}
\newcommand{\bear}[1]{\begin{eqnarray}\label{#1}}
\newcommand{\ear}{\end{eqnarray}}
\newcommand{\R}{{\mathbb R}}
\newcommand{\N}{{\mathbb N}}
\newcommand{\SL}{\mathop{\rm sl}\nolimits}
\newcommand{\rank}{\mathop{\rm rank}\nolimits}
\newcommand{\diag}{\mathop{\rm diag}\nolimits}
\newcommand{\sign}{\mathop{\rm sign}\nolimits}
\newcommand{\btd}{\bigtriangledown}
\newcommand{\btu}{\bigtriangleup}
\newcommand{\eps}{\varepsilon}
\newcommand{\tri}{\triangle}
\newcommand{\p}{\partial}
\newcommand{\fn}{\footnote}
\newcommand{\fnm}{\footnotemark}
\newcommand{\fnt}{\footnotetext}
 \def\nnn{\nonumber\\ &&\nqq}

\title{Black-brane solution for  $C_2$ algebra}

\begin{center} \large\bf
BLACK-BRANE SOLUTION FOR $C_2$ ALGEBRA
\\[15pt]
\normalsize\bf M.A. Grebeniuk
\fnm[1]\fnt[1]{mag@gravi.phys.msu.su},
V.D. Ivashchuk\fnm[2]\fnt[2]{ivas@rgs.phys.msu.su} \\[10pt]

\it Center for Gravitation and Fundamental Metrology, VNIIMS, 3/1
M. Ulyanovoy Str.,
Moscow 117313, Russia  and\\
Institute of Gravitation and Cosmology, People's Friendship
University \\
of Russia, Mikhlukho-Maklaya Str. 6,
\\ Moscow 117198, Russia \\
\vspace{7pt}
\normalsize\bf and S.-W. Kim\fnm[3]\fnt[3]{sungwon@mm.ewha.ac.kr}
 \\[10pt]
\it Department of Science Education and Basic Science Research Institute,
Ewha Women's University, Seoul 120-750, Korea

\end{center}

\vspace{15pt}


\begin{abstract}

Black $p$-brane  solutions  for a wide class of
intersection rules and  Ricci-flat ``internal'' spaces are considered.
They are defined up to moduli functions $H_s$ obeying non-linear
differential equations with  certain boundary conditions imposed.
A new solution with intersections corresponding to the Lie algebra $C_2$ is
obtained. The functions $H_1$ and $H_2$ for this solution are polynomials
of degree $3$ and $4$.

\end{abstract}

\bigskip

\hspace{1cm}*PACS numbers:
04.20.Jb, 04.50.+h, 04.70.Bw, 02.20.Sv, 02.30.Hq.

\newpage


\section{Introduction}

The paper is devoted to black-brane  solutions with
(next to) arbitrary intersections (see Section 2)
\cite{IMp1,IMp2,IMp3,IMtop}.
These black-brane solutions are governed by moduli functions $H_s =
H_s(R)$ obeying a set of second order non-linear differential
equations with some boundary relations imposed.

Some general features of the black-brane solutions were investigated earlier in \cite{Br2}.
More general spherically symmetric (and cosmological solutions)
were obtained in ref. \cite{IK} using the sigma-model approach and the
Lagrange representation from refs. \cite{IMC,IMJ}.

In \cite{IMp1} the following conjecture
was suggested (see Section 3): the moduli
functions $H_s$ are polynomials when intersection rules
correspond to semisimple Lie algebras. This conjecture was
confirmed by special  black-brane "block-orthogonal" solutions
considered earlier in \cite{Br,IMJ2,CIM} (subsect. 3.1). It was verified also for
$A_m$ and $C_m$ series of Lie algebras in \cite{IMp2,IMp3}. An analogue of
this conjecture for extremal black holes was considered earlier
in \cite{LMMP}.

An example of black brane solution corresponding to Lie algebra
$A_2 = sl(3)$ was considered earlier in refs. \cite{IMp1,IMp2},
where some dyonic configurations
(e.g. in $11$-dimensional supergravity) were
considered. The $A_2$-solution is governed by two polynomials
$H_1$ and $H_2$  of degree $2$.
The coefficients of polynomials and charges are functions
of some parameters $P_1$ and $P_2$  and have a rather simple
form (subsect. 3.1). It seems that $A_2$-solution is the only one
that may be obtained ``by hands'': the search  of  solutions for
other Lie algebras needs  computer calculations.

Here we present a first non-trivial result of such calculations, i.e.
a new solution corresponding to the Lie algebra $C_2$ governed by polynomials
of degree $3$ and $4$ (subsect. 3.3).
The coefficients of polynomials exhibit a non-trivial
dependence upon parameters $P_1$ and $P_2$,
namely: a non-trivial term $\Delta$, that is a
square root of a polynomial of degree $4$, appears.


\section{$p$-brane black hole solutions}

Consider a  model governed by the action
\ber{1.1}
S=\int d^Dx \sqrt{|g|}\biggl\{R[g]-h_{\alpha\beta}g^{MN}\p_M\varphi^\alpha
\p_N\varphi^\beta-\sum_{a\in\tri}\frac{\theta_a}{n_a!}
\exp[2\lambda_a(\varphi)](F^a)^2\biggr\}
\eer
where $g=g_{MN}(x)dx^M\otimes dx^N$ is a metric,
$\varphi=(\varphi^\alpha)\in\R^l$ is a vector of scalar fields,
$(h_{\alpha\beta})$ is a  constant symmetric
non-degenerate $l\times l$ matrix $(l\in \N)$,
$\theta_a=\pm1$,
\beq{1.2a}
F^a =    dA^a
=  \frac{1}{n_a!} F^a_{M_1 \ldots M_{n_a}}
dz^{M_1} \wedge \ldots \wedge dz^{M_{n_a}}
\eeq
is a $n_a$-form ($n_a\ge1$), $\lambda_a$ is a
1-form on $\R^l$: $\lambda_a(\varphi)=\lambda_{\alpha a}\varphi^\alpha$,
$a\in\tri$, $\alpha=1,\dots,l$.
In (\ref{1.1})
we denote $|g| =   |\det (g_{MN})|$, \\
$(F^a)^2_g  =
F^a_{M_1 \ldots M_{n_a}} F^a_{N_1 \ldots N_{n_a}}
g^{M_1 N_1} \ldots g^{M_{n_a} N_{n_a}}$,
$a \in \tri$. Here $\tri$ is some finite set.
In the models with one time all $\theta_a =  1$
when the signature of the metric is $(-1,+1, \ldots, +1)$.

Let us consider (black-brane) solutions
to field equations corresponding to the action
(\ref{1.1})  from \cite{IMp1,IMp2,IMp3}. These solutions are  defined on the manifold
\beq{1.2}
M =    (R_{0}, + \infty)
\times (M_1 = S^{d_1}) \times (M_2 = \R) \times  \ldots \times M_n,
\eeq
and have the following form
\bear{2.30}
g= \Bigl(\prod_{s \in S} H_s^{2 h_s d(I_s)/(D-2)} \Bigr)
\biggl\{ f^{-1} dR \otimes dR
+ R^2  d \Omega^2_{d_1}  \\ \nn
-  \Bigl(\prod_{s \in S} H_s^{-2 h_s} \Bigr)
f  dt \otimes dt
+ \sum_{i = 3}^{n} \Bigl(\prod_{s\in S}
  H_s^{-2 h_s \delta_{iI_s}} \Bigr) g^i  \biggr\},
\\  \label{2.31}
\exp(\varphi^\alpha)=
\prod_{s\in S} H_s^{h_s \chi_s \lambda_{a_s}^\alpha},
\\  \label{2.32a}
F^a= \sum_{s \in S} \delta^a_{a_s} {\cal F}^{s},
\ear
where $f =1 - 2\mu/R^{\bar d}$,
\beq{2.32}
{\cal F}^s= - \frac{Q_s}{R^{d_1}}
( \prod_{s' \in S}  H_{s'}^{- A_{s s'}}) dR \wedge\tau(I_s),
\quad s\in S_e,
\eeq
\beq{2.33}
{\cal F}^s= Q_s \tau(\bar I_s), \quad s\in S_m.
\eeq
Here $Q_s \neq 0$ ($s\in S$) are charges, $R_0 >0$,
$R_0^{\bar d} =2 \mu > 0$, $\bar d = d_1 -1$.
In  (\ref{2.30})
$g^i=g_{m_i n_i}^i(y_i) dy_i^{m_i}\otimes dy_i^{n_i}$
is a Ricci-flat  metric on $M_{i}$, $i=  3,\ldots,n$
and $\delta_{iI}=  \sum_{j\in I} \delta_{ij}$
is the indicator of $i$ belonging
to $I$: $\delta_{iI}=  1$ for $i\in I$ and $\delta_{iI}=  0$ otherwise.
Here  $g^2 = -dt \otimes dt$,
and $g^1 = d \Omega_{d_1}$ be a canonical metric
on unit sphere $M_1 =S^{d_1}$,

The  $p$-brane  set  $S$ is by definition
\beq{1.6}
S=  S_e \cup S_m, \quad
S_v=  \cup_{a\in\tri}\{a\}\times\{v\}\times\Omega_{a,v},
\eeq
$v=  e,m$ and $\Omega_{a,e}, \Omega_{a,m} \subset \Omega$,
where $\Omega =   \Omega(n)$  is the set of all non-empty
subsets of $\{ 2, \ldots,n \}$, i.e.
all $p$-branes do not ``live'' in  $M_1$.

Any $p$-brane index $s \in S$ has the form
$s =   (a_s,v_s, I_s)$,
where
$a_s \in \tri$, $v_s =  e,m$ and $I_s \in \Omega_{a_s,v_s}$.
The sets $S_e$ and $S_m$ define electric and magnetic $p$-branes,
correspondingly. In (\ref{2.31}) $\chi_s  =   +1, -1$
for $s \in S_e, S_m$, respectively. All $p$-branes
contain the time manifold $M_2 = \R$, i.e.
\beq{1.7a}
2 \in I_s, \qquad \forall s \in S.
\eeq

All  manifolds $M_{i}$, $i > 2$, are oriented and
connected and
\beq{1.12}
\tau_i  \equiv \sqrt{|g^i(y_i)|}
\ dy_i^{1} \wedge \ldots \wedge dy_i^{d_i},
\eeq
are volume $d_i$-forms, where $d_{i} =   {\rm dim} M_{i}$, $i =   1, \ldots, n$,
with $d_1 > 1$ and $d_2 = 1$. For any
 $I =   \{ i_1, \ldots, i_k \} \in \Omega$, $i_1 < \ldots < i_k$,
we denote
\beq{1.13}
\tau(I) \equiv \tau_{i_1}  \wedge \ldots \wedge \tau_{i_k},
\qquad
d(I)  =  \sum_{i \in I} d_i.
\eeq
The forms ${\cal F}^s$  correspond to electric
and magnetic $p$-branes for $s\in S_e, S_m$, respectively.
In (\ref{2.33}) we use the notation $\bar I= \{1,\ldots,n\}\setminus I$.

The parameters  $h_s$ appearing in the solution
satisfy the relations: $h_s = (B_{s s})^{-1}$, where
\beq{1.17}
B_{ss'} =
d(I_s\cap I_{s'})+\frac{d(I_s)d(I_{s'})}{2-D}+
\chi_s\chi_{s'}\lambda_{\alpha a_s}\lambda_{\beta a_{s'}}
h^{\alpha\beta},
\eeq
$s, s' \in S$, with $(h^{\alpha\beta})=(h_{\alpha\beta})^{-1}$
and $D =   1 + \sum_{i =   1}^{n} d_{i}$.
Here we assume that
$({\bf i})  B_{ss} \neq 0$,  $s \in S$,
and $({\bf ii}) {\rm det}(B_{s s'}) \neq 0$,
i.e. the matrix $(B_{ss'})$ is a non-degenerate one.

Let us consider the matrix
\beq{1.18}
(A_{ss'}) = \left( 2 B_{s s'}/B_{s' s'} \right).
\eeq
Here  some ordering in $S$ is assumed.

The functions $H_s = H_s(z) > 0$, $z = 2\mu/R^{\bar d} \in (0,1)$,
obey the equations
\beq{3.1}
 \frac{d}{dz} \left( \frac{(1-z)}{H_s} \frac{d H_s}{dz} \right) = B_s
\prod_{s' \in S}  H_{s'}^{- A_{s s'}},
\eeq
equipped with the boundary conditions
\bear{3.2a}
H_{s}(1 - 0) = H_{s0} \in (0, + \infty), \\
\label{3.2b}
H_{s}(+ 0) = 1,
\ear
$s \in S$. Here $B_{s} = B_{ss} \eps_s Q_s^2/(2 \bar d \mu)^2$
and
\beq{1.22}
\eps_s=(-\eps[g])^{(1-\chi_s)/2}\eps(I_s) \theta_{a_s},
\eeq
$s\in S$, $\eps[g]\equiv\sign\det(g_{MN})$. More explicitly eq.
(\ref{1.22}) reads: $\eps_s=\eps(I_s) \theta_{a_s}$ for
$v_s = e$ and $\eps_s=-\eps[g] \eps(I_s) \theta_{a_s}$, for
$v_s = m$.

Equations  (\ref{3.2a})  are equivalent to Toda-type
equations. The first boundary condition guarantees the existence
of a regular horizon at $R^{\bar{d}} =   2 \mu$. The second
condition (\ref{3.2b}) ensures an asymptotical
flatness (for $R \to +\infty$) of the $(2+d_1)$-dimensional
section of the metric.

Due to eqs. (\ref{2.32}) and  (\ref{2.33}), the dimension of
$p$-brane worldvolume $d(I_s)$ is defined by relations:
$d(I_s)=  n_{a_s}-1$, $d(I_s) = D- n_{a_s} -1$,
for $s \in S_e, S_m$, respectively.
For a $p$-brane we use a standard  notation: $p =   p_s =   d(I_s)-1$.

The solutions are valid if the following  restriction on the sets
$\Omega_{a,v}$ is imposed. This restriction guarantees the block-diagonal
structure of the stress-energy tensor.
We denote $w_1\equiv\{i|i\in \{2,\dots,n\},\quad d_i=1\}$ and
$n_1=|w_1|$ (i.e. $n_1$ is the number of 1-dimensional spaces among
$M_i$, $i=2,\dots,n$).  It is clear, that $2 \in w_1$.

{\bf Restriction.} {\em Let 1a) $n_1\le1$ or 1b) $n_1\ge2$ and for
any $a\in\tri$, $v\in\{e,m\}$, $i,j\in w_1$, $i \neq j$, there are no
$I,J\in\Omega_{a,v}$ such that $i \in I$, $j\in J$ and $I\setminus\{i\}=
J\setminus\{j\}$.}

This restriction is  satisfied in the non-composite case:
$|\Omega_{a,e}| + |\Omega_{a,m}| = 1$,
(i.e when there are no two  $p$-branes with the same color index $a$,
$a\in\tri$.)  The restriction forbids certain intersections of two
$p$-branes with the same color index for  $n_1 \geq 2$.


The solution under consideration describes a set of charged (by forms) overlapping
black $p$-branes "living" on submanifolds of $M_2 \times \dots \times M_n$.

\section{Examples of solutions}

\subsection{``Block-orthogonal'' solutions}

The simplest polynomial solutions occur in orthogonal
case \cite{CT,AIV,Oh,BIM,IMJ}, when
\beq{3.4}
B_{s s'} = 0,
\eeq
for  $s \neq s'$, $s, s' \in S$. In this case
$(A_{s s'}) = {\rm diag}(2,\ldots,2)$ is a Cartan matrix
for semisimple Lie algebra $A_1 \oplus  \ldots  \oplus  A_1$
and
\beq{3.5}
H_{s}(z) = 1 + P_s z,
\eeq
with $P_s \neq 0$, satisfying
\beq{3.5a}
P_s(P_s + 1) = - B_s,
\eeq
$s \in S$. For positive parameters $P_s > 0$  we get negative $B_s < 0$.

In \cite{Br,IMJ2,CIM} this solution
was generalized to the "block-orthogonal" case:
\beq{3.6}
S=S_1 \cup\dots\cup S_k, \qquad  S_i \cap S_j = \emptyset, \quad i \neq j,
\eeq
$S_i \ne \emptyset$, i.e. the set $S$ is a union of $k$ non-intersecting
(non-empty) subsets $S_1,\dots,S_k$,
and relation (\ref{3.4}) is satisfied
for all $s\in S_i$, $s'\in S_j$, $i\ne j$; $i,j=1,\dots,k$.
In this case eq. (\ref{3.5}) is modified as follows
\beq{3.8}
H_{s}(z) = (1 + P_s z)^{b^s},
\eeq
where
\beq{3.11}
b^s = 2 \sum_{s' \in S} A^{s s'},
\eeq
$(A^{s s'}) = (A_{s s'})^{-1}$ and parameters $P_s$ are coinciding inside
blocks, i.e. $P_s = P_{s'}$ for $s, s' \in S_i$, $i =1,\dots,k$.
Parameters $P_s \neq 0 $ satisfy the relations (\ref{3.5a})
and parameters $B_s$  are also  coinciding inside
blocks, i.e. $ B_s =  B_{s'}$ for $s, s' \in S_i$, $i =1,\dots,k$.

Let $(A_{s s'})$ be  a Cartan matrix  for a  finite-dimensional
semisimple Lie  algebra $\cal G$. In this case all powers in
(\ref{3.11})  are  natural numbers
coinciding with the components  of twice the dual Weyl
vector in the basis of simple roots \cite{FS},
and  hence, all functions $H_s$ are polynomials, $s \in S$.

{\bf Conjecture \cite{IMp1}.} {\em Let $(A_{s s'})$ be  a Cartan matrix
for a  semisimple finite-dimensional Lie algebra $\cal G$.
Then  the solution to eqs. (\ref{3.1})-(\ref{3.2b})
(if exists) is a polynomial
\beq{3.12}
H_{s}(z) = 1 + \sum_{k = 1}^{n_s} P_s^{(k)} z^k,
\eeq
where $P_s^{(k)}$ are constants,
$k = 1,\ldots, n_s$, the integers $n_s = b^s$ are
defined in (\ref{3.11}) and $P_s^{(n_s)} \neq 0$,  $s \in S$}.

This conjecture was verified for $A_n$ and $C_n$  series of Lie algebras
\cite{IMp2,IMp3}. In the extremal case
$\mu = + 0$ an a analogue of this conjecture was suggested previously in
\cite{LMMP}.

\subsection{Solution for $A_2$ algebra}

Here we present the polynomial
solution from \cite{IMp1,IMp2} corresponding to the
Lie algebra $A_2 = sl(3)$ with the Cartan
matrix
\beq{B.1a}
\left(A_{ss'}\right)=
\left( \begin{array}{*{6}{c}}
2&-1\\
-1&2\\
\end{array}
\right)\quad
\eeq

The moduli polynomials  read in this case as follows
\beq{4.1}
H_{s} = 1 + P_s z + P_s^{(2)} z^{2},
\eeq
where $P_s= P_s^{(1)}$ and $P_s^{(2)} \neq 0$ are constants and
\bear{4.5}
 P_s^{(2)} = \frac{ P_s P_{s +1} (P_s + 1 )}{2 (P_1 +P_2 + 2)},
 \\ \label{4.6}
B_s = - \frac{ P_s (P_s + 1 )(P_s + 2 )}{P_1 +P_2 + 2},
\ear
$s = 1,2$.  Here $P_1 +P_2 + 2 \neq 0$.

In the $A_2$-case the solution is described by relations
(\ref{2.30})-(\ref{2.33}) with $S = \{s_1,s_2\}$ and
intersection rules following from (\ref{1.17}), (\ref{1.18})
and (\ref{B.1a})
\bear{1.40a}
d(I_{s_1} \cap I_{s_2})= \frac{d(I_{s_1})d(I_{s_2})}{D-2}-
\chi_{s_1} \chi_{s_2} \lambda_{a_{s_1}}\cdot\lambda_{a_{s_2}}
- \frac12 K,
\\ \label{1.40b}
d(I_{s_i}) - \frac{(d(I_{s_i}))^2}{D-2}+
\lambda_{a_{s_i}}\cdot\lambda_{a_{s_i}} = K,
\ear
where
$K \neq 0$  and functions $H_{s_i} = H_i$
are defined by relations
(\ref{4.1})-(\ref{4.6}) with $z = 2\mu R^{-\bar d}$,
$i =1,2$. Here and in what follows $\lambda\cdot\lambda^{'}=
\lambda_{\alpha}\lambda_{\beta }^{'}h^{\alpha\beta}$.

\subsection{Solutions for $C_2$ algebra}

Now we present the solution related to the Lie algebra $C_2 =
so(5)$ with the Cartan matrix
\beq{B.1c}
\left(A_{ss'}\right)= \left( \begin{array}{*{6}{c}}
2&-1\\
-2&2\\
\end{array}
\right). \quad
\eeq

According to {\bf Conjecture} we  seek the
solution to eqs. (\ref{3.1})-(\ref{3.2b}) in the following form
\bear{5.1}
 H_1(z)=1+P_1z+P_1^{(2)}z^2+P_1^{(3)}z^3, \\
 \label{5.2}
H_2(z)=1+P_2z+P_2^{(2)}z^2+P_2^{(3)}z^3+P_2^{(4)}z^4,
 \ear
where $P_s= P_s^{(1)}$ and $P_s^{(k)}$ are constants, $s = 1,2$.

Here we outline the result.
For $B_s$-parameters we get the following relations
 \bear{5.3}
 2B_1=-\Delta + (2P_1+ 3)(2+P_2),
 \\ \label{5.4}
B_2=\Delta -  2 -2P_1 (P_1 + 3)-(2+P_2)^2,
\ear
and for parameters $P_s^{(k)}$ we obtain
\bear{5.5}
4 P_1^{(2)}= 6+3P_2-\Delta +2P_1(3+P_1+P_2), \qquad
\\  \label{5.6}
12P_1^{(3)}=
- \Delta (2+P_1+P_2)+
12+18P_1+2P_1^3+3P_2(4+P_2)+
\\ \nn
 2 P_1^2(5+P_2)+P_1P_2(11+2P_2),
\\  \label{5.7}
2 P_2^{(2)}= -6- 2P_1(3+P_1)-3P_2+\Delta, \qquad
\\  \label{5.8}
6P_2^{(3)}= \Delta (2+2P_1+P_2)-12-24P_1-4P_1^3-3P_2(4+P_2)
\\ \nn
- 2P_1P_2(7+P_2)-2P_1^2(8+P_2),
\\  \label{5.9}
24P_2^{(4)}= \Delta [2P_1^2+(3+P_2)(2+2P_1+P_2)]
\\ \nn
-4P_1^4-3(2+P_2)^2(3+P_2)-2P_1(3+P_2)^2(4+P_2)-
\\ \nn
4P_1^3(6+P_2)-P_1^2(60+30P_2+4P_2^2),
\ear
where
 \beq{5.10}
\Delta = \sqrt{4\left(3+P_1(3+P_1)\right)^2+(3+2P_1)^2P_2(4+P_2)}.
\eeq
It may be verified that $B_1 < 0$ and $B_2 < 0$ for $P_1 > 0$, $P_2 > 0$.

The $C_2$ black-brane solution is described by relations
(\ref{2.30})-(\ref{2.33}), with $S = \{s_1,s_2\}$, and intersection
rules following from (\ref{1.17}), (\ref{1.18}) and (\ref{B.1c})
\bear{5.11a}
d(I_{s_1} \cap I_{s_2})=
\frac{d(I_{s_1})d(I_{s_2})}{D-2}- \chi_{s_1} \chi_{s_2}
\lambda_{a_{s_1}}\cdot\lambda_{a_{s_2}} -  K,
\\ \label{5.11b}
d(I_{s_1}) - \frac{(d(I_{s_1}))^2}{D-2}+
\lambda_{a_{s_1}}\cdot\lambda_{a_{s_1}} = K,
\\ \label{5.12c}
d(I_{s_2}) - \frac{(d(I_{s_2}))^2}{D-2}+
\lambda_{a_{s_2}}\cdot\lambda_{a_{s_2}} = 2K,
 \ear
where $K \neq 0$  and functions $H_{s_i} = H_i$ are defined
by relations (\ref{5.1}) and (\ref{5.2})
 with $z = 2\mu R^{-\bar d}$, $i =1,2$.

A simple test for verification
of the relations for $H_s$-functions is to put
$P_1 = 3P$ and $P_2 = 4P$ with $P > 0$.
In this case we a get a special ``block-orthogonal''
solution
 \beq{5.1b}
  H_1(z)= (1+ P z)^3, \qquad  H_2(z)= (1+ P z)^4,
  \eeq
in agreement with the relations (\ref{3.8}) and (\ref{3.11}).

\section{Conclusions}

 In this paper we presented a new black-brane solution with two
 branes and  intersection rule corresponding to the Lie algebra
 $C_2$. The solutions is governed by polynomials of degree $3$ and $4$
 and the coefficients of polynomials exhibit a non-trivial
 dependence upon the parameters $P_1$ and $P_2$,
 due to appearance of the non-trivial term $\Delta$ (see (\ref{5.10})).
 The $C_2$-solution differs drastically from the $A_2$-one,
 that has a rather simple analytical
 structure.  This means that the polynomial solutions  for other
 $A_n$-algebras ($n > 2$) should be also non-trivial.
 Indeed, the $C_2$-solution may be extended to a special
 $A_3$-one, if we impose a restriction on the moduli polynomials
 $H_1$, $H_2$, $H_3$  of the following form: $H_1= H_3$. (Thus, here
 we obtained  by product a special $A_3$ black-brane solution.)
 It should be noted  that $A,D,E$ (or simply laced) Lie algebras are
 of much interest, since they  appear for $p$-brane intersection rules
 in supergravitational models \cite{IMJ}. Another topics of
 interest are related to black-brane thermodynamics
 (e.g. relations for the entropy,
 the Hawking temperature etc) and analysis of post-Newtonian effects.
 On this way one may expect to clarify the appearance of the
 function $\Delta$
 in the solution. But the main (mathematical) problem
 here is to find the polynomial solutions for all Lie algebras. This
 problem seems to be a very difficult one and may be of interest
 for mathematicians dealing with polynomials
 (e.g. appearing in non-linear ordinary and partial differential
 equations), Lie algebras, number theory etc.

\begin{center}
{\bf Acknowledgments}
\end{center}

This work was supported in part by the Russian Ministry for
Science and Technology, Russian Foundation for Basic Research,
project SEE,  by grant No. R01-2000-00015
>from the Korea Science and Engineering Foundation and the MOST
through National Research Program (00-B-WB-06-A-04) for
Women's University. V.D.I. thanks the colleagues from Ewha Women's
University for kind hospitality.
The results of the work were  reported at
the workshop "International Camp of Gravitation"
(Sept. 8-9, 2001, Institute of Wind and Water, Gapyong, Korea)
and at Fifth International conference on Gravitation
and Astrophysics of Asia-Pacific countries.



\begin{thebibliography}{29}

\bibitem{IMp1}
V.D. Ivashchuk and V.N. Melnikov, P-brane black holes for general
intersections, {\it Gravitation \& Cosmology} {\bf 5}, 313-318
(1999); gr-qc/0002085.

\bibitem{IMp2}
V.D. Ivashchuk and V.N. Melnikov, Black hole p-brane solutions for
general intersection rules. {\it Gravitation \& Cosmology} {\bf 6},
27-40 (2000); hep-th/9910041.

\bibitem{IMp3}
V.D. Ivashchuk and V.N. Melnikov, Toda  p-brane black holes and
polynomials related to Lie algebras. {\it Class. Quantum Grav.}
{\bf 17}, 2073-2092 (2000); math-ph/0002048.

\bibitem{IMtop}
V.D. Ivashchuk and V.N. Melnikov, Exact solutions in
multidimensional gravity with antisymmetric forms, topical review,
{\it Class. Quantum Grav.}  {\bf 18}, R1-R66 (2001);
hep-th/0110274.

\bibitem{IK}
V.D.Ivashchuk and S.-W. Kim, Solutions with intersecting p-branes
related to Toda chains, {\it J. Math. Phys.} {\bf 41}, 444-460
(2000); hep-th/9907019.

\bibitem{IMC}
V.D. Ivashchuk and V.N. Melnikov, Sigma-model for the
generalized  composite p-branes, {\it Class. Quantum Grav.} {\bf
14}, 3001-3029 (1997); {\it Corrigenda} {\bf 15 }, 3941
(1998); hep-th/9705036.

\bibitem{IMKor}
V.D.Ivashchuk and V.N.Melnikov, Exact solutions in sigma-model
with p-branes, {\it J. Korean Phys. Soc.}
{\bf 35}, S638-S648 (1999).

\bibitem{CT}
M. Cvetic and A. Tseytlin, Non-extreme black holes from
non-extreme intersecting $M$-branes, {\it Nucl. Phys.} {\bf B
478}, 181-198 (1996); hep-th/9606033.

\bibitem{AIV}
I.Ya. Aref'eva, M.G. Ivanov and I.V. Volovich, Non-extremal
intersecting p-branes in various dimensions,  {\it Phys. Lett. }
{\bf B 406}, 44-48 (1997); hep-th/9702079.

\bibitem{Oh}
N. Ohta, Intersection rules for non-extreme p-branes, {\it Phys.
Lett. }  {\bf B 403}, 218-224 (1997); hep-th/9702164.

\bibitem{LMMP}
H. L\"u, J. Maharana, S. Mukherji  and C.N. Pope, Cosmological
solutions, p-branes and the Wheeler-DeWitt equation, {\it Phys.
Rev. } {\bf D 57}, 2219-2229 (1998); hep-th/9707182.

\bibitem{BIM}
K.A. Bronnikov, V.D. Ivashchuk and V.N. Melnikov, The
Reissner-Nordstr\"om problem for intersecting electric and
magnetic  p-branes,  {\it Gravitation \& Cosmology} {\bf 3},
203-212 (1997); gr-qc/9710054.

\bibitem{IMJ}
V.D. Ivashchuk and V.N. Melnikov, Multidimensional classical and
quantum cosmology with intersecting $p$-branes, {\it J. Math.
Phys.} {\bf 39}, 2866-2889 (1998); hep-th/9708157.

\bibitem{Br}
K.A. Bronnikov, Block-orthogonal brane systems, black holes and
wormholes, {\it Gravitation \& Cosmology} {\bf 4}, 49-56 (1998);
hep-th/9710207.

\bibitem{Br2}
K.A. Bronnikov, Gravitating brane systems: some general theorems,
{\it J. Math. Phys.} {\bf 40}, 2, 924-938 (1999); gr-qc/9806102.

\bibitem{IMJ2}
V.D. Ivashchuk and V.N. Melnikov. Multidimensional cosmological and
spherically symmetric solutions with intersecting p-branes.
In Lecture Notes in Physics, Vol. 537,
{\it Mathematical and Quantum Aspects of Relativity and Cosmology},
Proceedings of the Second Samos Meeting on Cosmology, Geometry and
Relativity held at Pythagoreon, Samos, Greece, 1998,
eds. S. Cotsakis, G.W. Gibbons (Springer, Berlin, 2000);
gr-qc/9901001.

\bibitem{IMJ1}
V.D.Ivashchuk and V.N.Melnikov, Cosmological and spherically
symmetric solutions with intersecting p-branes. {\it J. Math. Phys.}
{\bf 40}, 6558-6576 (1999).

\bibitem{CIM}
S. Cotsakis, V.D. Ivashchuk and V.N. Melnikov,
P-branes black holes and post-Newtonian approximation,
{\it Gravitation \& Cosmology} {\bf 5},
52-57 (1999); gr-qc/9902148.

\bibitem{FS}
J. Fuchs and C. Schweigert, {\it Symmetries, Lie algebras and
representations. A graduate course for physicists}
(Cambridge University Press, Cambridge, 1997).

\end{thebibliography}
\end{document}